\documentclass[prb,twocolumn,showpacs,amsmath,amssymb,superscriptaddress]{revtex4}
\usepackage{graphicx}
\usepackage{dcolumn}
\usepackage[sort&compress]{natbib}
\usepackage{subfigure}
\usepackage{ifpdf}
\usepackage{bm}
\ifpdf
\usepackage[pdftex,
        colorlinks=true,
        pdftitle={},
        pdfauthor={S. Salem-Sugui Jr.},
        pdfsubject={phase coherence},
        pdfkeywords={IF, UFRJ, Rio de Janeiro Brazil},
        baseurl={http://www.if.ufrj.br},
        pdfpagemode=UseNone,
        bookmarksopen=true
        pdfoutput=1
        ]{hyperref}
\fi

\begin{document}
\title{Study of diamagnetic  fluctuations in YBa$_2$Cu$_3$O$_{7-x}$  and Bi$_2$Sr$_2$CaCu$_2$O$_{8+x}$: Possible observation of phase correlations persisting above $T_c$}
\author{S. Salem-Sugui Jr.}
\affiliation{Instituto de Fisica, Universidade Federal do Rio de Janeiro,
21941-972, Rio de Janeiro, RJ, Brazil}
\author{J. Mosqueira}
\affiliation{LBTS,Universidade de Santiago de Compostela, E-15782, Spain}
\author{A. D. Alvarenga}
\affiliation{Instituto Nacional de Metrologia Normaliza\c{c}\~ao e
Qualidade Industrial, 25250-020 Duque de Caxias, RJ, Brazil.}
 \date{\today}
\begin{abstract}

We report on isofield curves of $\sqrt{-M}$ vs. $T$, where $M$ is the reversible magnetization, of YBa$_2$Cu$_3$O$_{6.95}$,  YBa$_2$Cu$_3$O$_{6.65}$, and Bi$_2$Sr$_2$CaCu$_2$O$_{8+x}$ with the magnetic field, $H$, applied parallel to the $c$-axis of the samples (and also parallel to the $ab$ planes for YBaCuO). For temperatures close to the critical temperature, $T_c$, the quantity $\sqrt{-M}$ is proportional to the order parameter amplitude $|\psi|$. Curves of $\sqrt{-M}$ vs. $T$ allowed to study the asymptotic behavior of the form $(T_a-T)^m$ of $|\psi|$ near $T_c$, as a function of field. Results for the studied samples produced values of $T_a(H)$ lying above $T_c$, suggesting that the magnetic field gradually allows to probe a region of temperatures where phase correlations persist above $T_c$. The study performed here in YBaCuO samples allowed to study how phase correlations evolve with doping in the pseudo-gap region of YBaCuO. $\sqrt{-M}$ vs. $T$ curves for all samples show a rather large  amplitude fluctuation with no phase correlation extending well above $T_a(H)$ which is interpreted in terms of a Gaussian Ginzburg-Landau approach with a \textit{total-energy} cutoff in the fluctuation spectrum. Resulting values for the exponent $m$ found for all samples, $0.5 < m < 0.7$, are interpreted as due to phase fluctuations of the $d$-wave pairing symmetry of the order parameter in the $ab$ planes. 
\end{abstract}
\maketitle 

\section{Introduction}
 It is well know that underdoped high-$T_c$ superconductors exhibit superconducting fluctuations extending well above the superconducting temperature transition for zero magnetic field, $T_c$, in the pseudo-gap region \cite{timusk,lee,tesa2}. Such diamagnetic fluctuations are thought to be due to amplitude fluctuations with no phase coherence \cite{wang1,wang2,jesusprl} where $T_c$ represents the temperature at which phase coherence is lost \cite{emery}. Based on Ref.~\cite{emery} superconductors with a small superfluid density, as underdoped high-$T_c$ superconductors, have a small phase stiffness and, as a consequence, phase fluctuations are important in the vicinity of the superconducting temperature.  More recently, a theoretical work which explicitly considers the effect of phase fluctuations on the phase diagram of a high-$T_c$ superconductor with a $d$-wave order parameter symmetry was presented in Ref.~\cite{beck}. In this work a phase diagram is obtained for YBaCuO as a function of doping showing a new line lying above $T_c$ (there denominated $T_\phi$ line)   representing the temperature at which phase coherence is lost. This $T_\phi$ line represents a limited temperature region above $T_c$ where phase correlations survive. The importance of phase fluctuations in $d$-wave coupling superconductors is also evidenced in another theoretical work presented in Ref.~[\onlinecite{colocar}]. These theoretical works of Refs.~[\onlinecite{emery,beck,colocar}] do not consider an applied magnetic field, but most of  the superconducting phenomena   observed above $T_c$ in these underdoped systems depend on the application of a magnetic field \cite{lee,wang1,wang2,rosenstein,said} and the common thought regarding fluctuations above $T_c$ is the existence of a state with vortices with no phase coherence \cite{wang1,wang2,jesusprl}. In a recent work \cite{op}, which include some of us, it was shown that the asymptotic behavior of isofield $\sqrt{-M}$ vs. $T$ curves near $T_c$ in two deoxygenated YBaCuO crystals supports the existence of a superconducting state with phase coherence persisting above $T_c$ as predicted in Ref.~[\onlinecite{beck}] while results for an optimally doped YBaCuO crystal indicated that phase coherence is controlled by the mean field $T_c(H)$ in disagreement with theory predictions \cite{emery, beck}. 
 
In this work, we address the issue of phase correlations above $T_c$ in the high-$T_c$ compounds YBa$_2$Cu$_3$O$_{6.65}$, YBa$_2$Cu$_3$O$_{6.95}$, and Bi$_2$Sr$_2$CaCu$_2$O$_{8+x}$ (Bi2212). Deoxygenated YBaCuO and Bi2212 systems exhibit the pseudo-gap phase \cite{timusk,lee} and phase fluctuations may play an important role in the vicinity of $T_c$ \cite{emery}. Isofield magnetization curves, M vs. T,  were obtained for YBaCuO crystals with the magnetic field, H, applied parallel and perpendicular to the $c$-axis while for Bi2212 we used previous obtained magnetization curves for $H \parallel c$-axis \cite{said}.  The YBaCuO crystals data complement a previous study performed in the vicinity of $T_c$ on two deoxygenated samples \cite{op} and verify (by studying another samples with $T_c$=92 K) a result obtained in Ref.[\onlinecite{op}] for an optimally doped sample, where no phase correlations were observed above $T_c$.  The work was motivated on the possibility of experimentally obtain a qualitative diagram showing the evolution  of phase correlations with doping in YBaCuO. Other motivations were to obtain the extension in temperatures above $T_c$ where phase correlations survive in Bi2212 and to study the large amplitude fluctuation observed well above $T_c$ on the studied systems. We study isofield curves of  $\sqrt{-M}$~vs. $T$ instead M vs. $T$ curves, since the quantity $\sqrt{-M}$ is directly proportional to the amplitude of the order parameter $|\psi|$ near $T_c$ \cite{deGennes}. The analysis is performed in two distinct regions of the isofield $\sqrt{-M}$vs.$T$ curves. Below $T_c(H)$ the curves show an asymptotic behavior of the form $(T_a-T)^m$, where $T_a(H)$ is an apparent transition temperature representing the onset of phase correlations, and $m$ is a fitting exponent. This analysis was motivated on the results of Ref.~[\onlinecite{kwon}] which show that phase fluctuations have an effect on the superfluid density of states reducing the gap in the vicinity of $T_c$. The region above $T_c(H)$, dominated by amplitude fluctuations, is analysed following a Gaussian GL approach developed in Refs.~[\onlinecite{EPLVidal,PC_Carlos}].
We also present  low isofield data for Bi2212 to demonstrate a similar crossing point  to data at higher fields that suggests further theoretical work.

The paper is organized as follows: In Section II we present the experimental details, and in Section III the results and discussion: in subsection III.A, we discuss the crossing point of the Bi2212 curves, in III.B the asymptotic behavior of the curves near $T_c$, in III.C the effect of amplitude fluctuations above $T_c$, and finally in III.D we present the phase diagram for the superconducting phase fluctuations in YBaCuO.

\section{Experimental}

We  measured an optimally doped YBa$_2$Cu$_3$O$_{6.95}$ single crystal with $T_c=92$~K and  a deoxygenated  YBa$_2$Cu$_3$O$_{6.65}$ crystal with $T_c=61.5$~K. YBaCuO crystals used here were grow at Argonne National Laboratory by Boyd Veal \cite{veal}.  To fix the oxygen stoichiometry, the sample was held for 89 hours at 520 $^{\circ}$C in a flowing gas atmosphere consisting of 3.45 $\%$$O_2$ in $N_2$.  The gas mixture was continuously monitored during the heat treatment with a commercial zirconia cell. 
The sample was then quenched in a liquid nitrogen bath.  This procedure fixes the oxygen stoichiometry at x=6.65, as discussed in Ref.~[\onlinecite{veal}]. The 92 K sample was treated for 100 hr at 480 $^{\circ}$C plus 190 hr at 400 $^{\circ}$C in flowing O2. The studied YBaCuO samples exhibited sharp, fully developed transitions, with a transition width $\Delta T_{c}\simeq1$ K. Magnetization curves obtained for YBaCuO samples studied here as obtained as a function of temperature for  $H\simeq 5$ Oe are presented in the
inset of Fig.1a. We also include in this inset of Fig. 1a low field curves obtained for deoxygenated YBaCuO crystals of Rev.~[\onlinecite{op}] which results are used in this work. The temperature at the middle value of the transition is taken as $T_c$. Dimensions of the samples used here are of the order $1\times1\times0.2$ mm (the $c$-axis being perpendicular to the largest plane of the samples) and their mass about 1 mg. For Bi$_2$Sr$_2$CaCu$_2$O$_{8+x}$ we used magnetization data recently obtained in a single crystal with  $T_c= 93$~K for $H\parallel c$-axis \cite{said}. Magnetization measurements (isofield $M$ vs. $T$ curves) were obtained by using a commercial magnetometer (Quantum Design) based on the superconducting quantum interference device (SQUID). All data were obtained after cooling the sample from a temperature above $T_c$ in the presence of the earth magnetic field, to a desired temperature below $T_c$. After that, a magnetic field was applied reaching the desired value without overshooting, and data were obtained by heating the sample at fixed increments of temperature up to a temperature well above $T_c$. Magnetization data for YBaCuO crystals were obtained for both field directions, $H\parallel ab$-planes and $H\parallel c$-axis. Magnetization data was obtained up to 150 K for sample with $T_c=92$~K (for $H\parallel c$-axis only)  and up to 110 K  for sample with $T_c=61.5$~K. Figure 1 shows selected isofield magnetization curves as obtained for both samples with $H \parallel c-axis$. As shown in Fig. 1a and 1b background magnetization has a typical diamagnetic signal which increases with field. The same trend is followed for data with $H \parallel ab$-planes (not shown). The background magnetization due to the normal electrons, $M_{\rm back}$, was obtained and removed for each data set by selecting data well above $T_c$ and fitting to the form $M_{\rm back}=c(H)/T+a(H)$.
\begin{figure}[t]
\includegraphics[width=\linewidth]{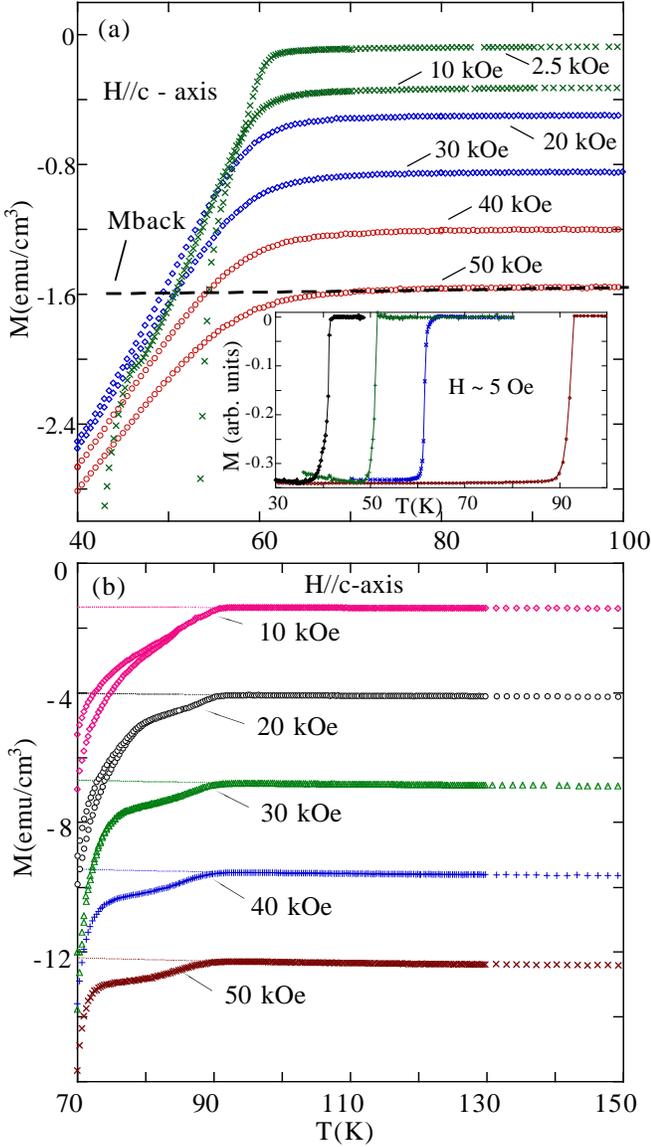}
\caption{Selected isofield M vs. T  curves as obtained with $H\parallel c$ for: (a)  the deoxygenated YBaCuO ($T_c$=61.5~K), where $M_{back}=-0.043-0.0293*H$ with H in kOe. (b) the optimally-doped YBaCuO ($T_c$=92~K) where $M_{\rm back}=c(H)/T+a(H)$ with $c(H)=(2.6+2.0*H)*10^{-4}$ and $a(H)=(44-8.9*H)*10^{-5}$ with H in kOe. Straight lines appearing below $T_c$ on the curves of Fig. 1b represent extrapolation of the background magnetization. The inset of Fig. 1a displays the superconducting transitions of the YBaCuO crystals}
\label{fig1}
\end{figure}

\begin{figure}[t]
\includegraphics[width=\linewidth]{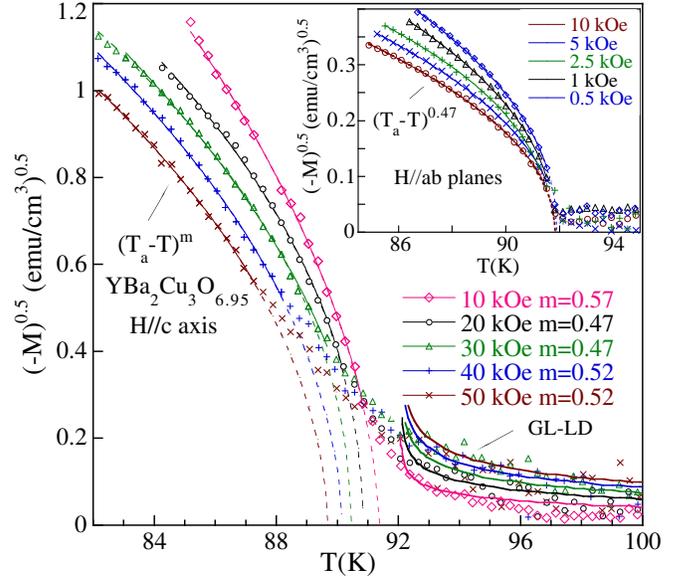}
\caption{Isofield curves of $\sqrt{-M}$ vs. T for the optimally-doped YBaCuO ($T_c$=92~K) with $H
\parallel c$. Dashed lines below $T_c$ show the extrapolation of the fittings to $T_a(H)$. Solid lines appearing above 92~K in the main figure represent fittings obtained from Eq.~(\ref{Schmidt3D}). 
Inset: Optimally-doped YBaCuO with $H\parallel ab$.}
\label{fig2}
\end{figure}

\begin{figure}[t]
\includegraphics[width=\linewidth]{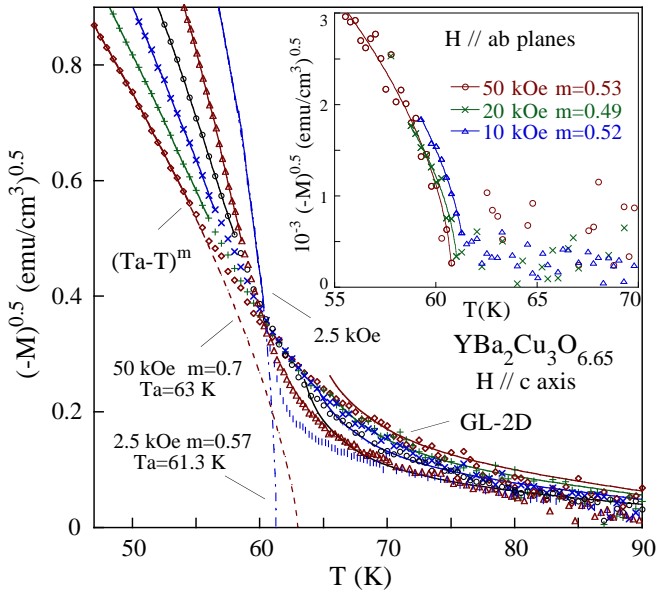}
\caption{Selected Isofield curves of $\sqrt{-M}$ vs.\ $T$ for the deoxygenated YBaCuO ($T_c$=61.5~K) with $H\parallel c$ for fields H=2.5, 10, 20, 30, 40, 50 kOe. Dashed  lines below 65~K shown the extrapolation of the fittings to $T_a(H)$ for fields $H$=2.5 and 50 kOe. Thicker dashed lines above $T_c$ represent fittings of Eq.~(\ref{Schmidt2D}) to the data with $H$=20, 30, 40 and 50 kOe. Inset: Deoxygenated YBaCuO with $H\parallel ab$.}
\label{fig3}
\end{figure} 

\begin{figure}[t]
\includegraphics[width=\linewidth]{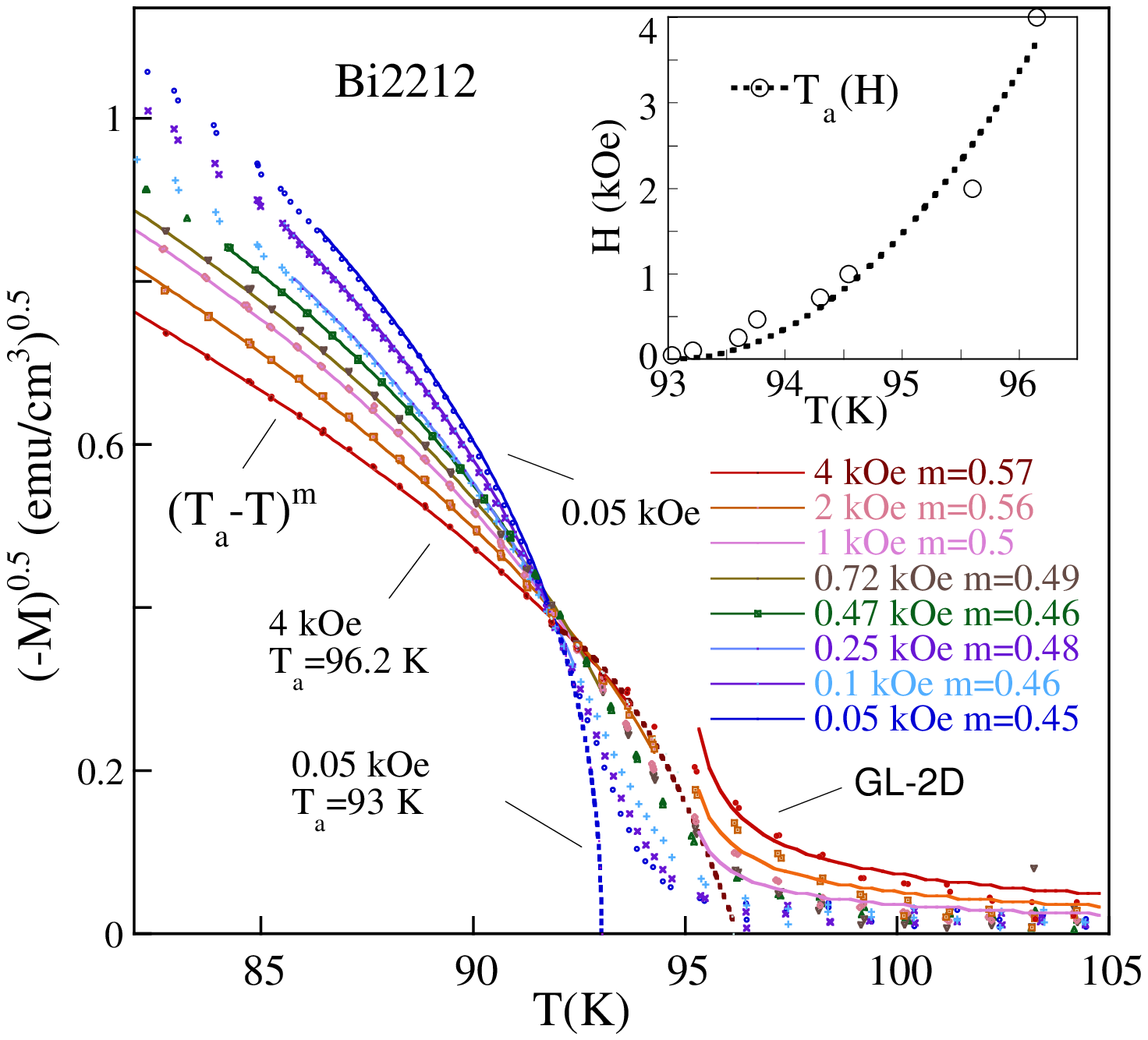}
\caption{Isofield curves of $\sqrt{-M}$ vs.\ $T$ for Bi2212 with $H\parallel c$. Dashed  lines  show the extrapolation of the fittings to $T_a(H)$ for fields $H$=0.05 and 4 kOe. Solid lines above $T_c$ represent fittings of Eq.~(\ref{Schmidt2D}) to the data with $H$=1, 2 and 4 kOe. Inset: Resulting $H$ versus $T_a(H)$ diagram, where dashed line is only a guidance to the eyes.}
\label{fig4}
\end{figure} 

\section{Results and discussion}

\subsection{Crossing point of the Bi2212 curves}

The isothermal magnetization curves of the studied systems are plotted as $\sqrt{-M}$ vs. $T$ and presented in Fig.~\ref{fig2} for YBa$_2$Cu$_3$O$_{6.95}$, Fig.~\ref{fig3} for YBa$_2$Cu$_3$O$_{6.65}$,  and Fig.~\ref{fig4} for Bi$_2$Sr$_2$CaCu$_2$O$_{8+x}$. The common fact among the curves obtained with field applied parallel to the $c$-axis, is a crossing point appearing below $T_c$ which defines a field independent magnetization for the corresponding set of curves. This crossing point, which occur for two-dimensional \cite{bula,tesa,rosenstein2} and three-dimensional \cite{zac1,rosenstein2} systems, are explained in terms of vortex fluctuations. We will not discuss this feature here, but we shall call the attention to the crossing point observed for Bi$_2$Sr$_2$CaCu$_2$O$_{8+x}$ in Fig.~\ref{fig4}. An inspection of this figure shows that the crossing point is formed by curves obtained for magnetic fields from 50~Oe to 4~kOe. Previous works \cite{tesa,bula} showing the same crossing point of magnetization curves  in Bi2212 were obtained with fields in the kOe range (the lowest field in Ref.~[\onlinecite{bula}] is 1~kOe, while the lowest field here is 0.05~kOe). The fact that very low field curves also cross the same point defined by intermediate-high fields may deserve further considerations since the theory predicting the existence of this crossing point \cite{bula,tesa,rosenstein2} is worked out in the lowest-Landau-level approximation. On the other hand the crossing point for a quasi-two-dimensional system is obtained in Ref.~[\onlinecite{tesa}] from a critical fluctuation theory. Then, the fact that isothermal curves for much lower fields cross the same point in Fig.~\ref{fig4} suggests that this system, which is known to have a very large anisotropy ($\gamma \geq 100$) may have a two-dimensional characteristic in the vicinity of $T_c$ even for very low fields \cite{klemm}, as observed in deoxygenated YBaCuO \cite{alvarenga}. To check out this possibility, we estimate the value of the parameter defined in Ref.~[\onlinecite{ger}] $r=8[\xi_{ab}(0)/\pi\gamma s]^2$ for Bi2212 with $T_c$= 93~K, where $\gamma$ is the anisotropy parameter which is assumed to be of the order of 100, $\xi_{ab}^2(0)= \phi_{0}/2\pi T_c |dH_{c2}/dT|_{T_c}$ is the in-plane Ginzburg-Landau 
coherence length extrapolated to $T$=0~K, $\phi_{0}$ is the flux quantum, and $s\sim15$ \r{A} \cite{bula} the CuO$_2$ layers periodicity. The parameter $r$ above  coincides at $T=T_c$ for dirty systems with the same parameter first defined in Ref.~[\onlinecite{klemm}] which carries 
important information about the dimensionality of the system. The calculated 
value is $r \simeq 6\times10^{-5}$  where we used $\gamma = 100$ and $\xi_{ab}(0)=13$ \r{A} (which results from the analysis of the fluctuation diamagnetism above $T_c$, see below). A plot of the reduced magnetic field $h\equiv H/H_{c2}(0)$ vs. $r$ is presented in Figure 9 of Ref.~[\onlinecite{klemm}], which helps to identify the dimensionality of fluctuations in a given system when the value of $r$ is know, and also predicts whether a given system can exhibit a field-induced-dimensionality crossover (3D toward 2D) in the vicinity of $T_c$. An inspection on this figure suggests that a system with $r \simeq6\times10^{-5}$ may enter the superconducting region exhibiting two-dimensional fluctuations for very low fields, which explains the crossing point observed here in Bi2212.

\subsection{Asymptotic behavior of the $\sqrt{-M}$ vs. $T$ curves near $T_c$}

Near the upper critical field $H_{c2} $ \cite{abrikosov} the magnetic induction $B$ obtained from the Landau-Ginzburg equation is given by \cite{deGennes}
\begin{equation} 
B=H-\frac{4\pi e \hbar}{mc}|\psi|^2 
\end{equation}
where $\psi$ is the order parameter and $B$ is the magnetic
induction. The magnetization $M=(B-H)/4\pi$ is
\begin{equation}  
M=-\frac{e \hbar}{mc}|\psi|^2 
\end{equation}
and  follows that $\sqrt{-M}$ is directly proportional to the average amplitude of the order parameter.  Then,  the quantity $\sqrt{-M}$ can be expressed near $T_c$ in the form $\sqrt{-M} \propto [T_c(H)-T]^m$ where $T_c(H)$ is the  mean field transition temperature. The mean field value of the critical exponent is $m=1/2$ for a $s$-wave BCS superconductor \cite{deGennes} and for a $d$-wave superconductor within a Ginzburg-Landau theory \cite{xu}. 

Here we study the presence of a phase-mediated transition by fitting to each curve the general scaling form $\sqrt{-M} \propto (T_a-T)^m$, where $T_a(H)$ is an apparent transition temperature representing the onset of phase correlations, and $m$ is a fitting exponent. As already mentioned this fitting is motivated by the results of Ref.~[\onlinecite{kwon}] which show that phase fluctuations of a $d$-wave order parameter have an effect on the superfluid density of states reducing the gap in the vicinity of $T_c$. The authors of Ref.~[\onlinecite{kwon}] show that this is a direct result of the $d$-wave order parameter symmetry which presents a node and anti-node where the effects of phase and amplitude fluctuations are distinct.  This result implies that the temperature dependence of the gap near $T_c$ is controlled by phase fluctuations. It is shown in Ref.~[\onlinecite{kwon}] that the overall effect of phase fluctuations produces a change in the shape (temperature dependence) of the gap increasing the value of the exponent $m$ \cite{op}. The phase mediated behavior of interest here is within a region where $\sqrt{-M}$ should follow the typical asymptotic behavior of the superconducting gap near $T_c$, which in our curves occur below the region where exist a large amplitude fluctuation extending above $T_c$. The inflection point occurring very close to $T_c$, which is clearly visible in each curve, marks a crossover between high temperature amplitude fluctuations and lower temperature phase fluctuations. The inflection point for each curve occurs very close to  the crossing point. The fitting is then performed in each curve in a temperature window delimited at higher temperatures by the inflection point occurring near but below $T_c$, and at lower temperatures by the region where the  magnetization show a linear dependence with temperature. The lower temperature cutoff lies within the region where the Abrikosov approximation above discussed is valid.  The fitting is usually performed in a temperature window of 5-10 K. Results for the fitted values of  $T_a(H)$ and $m$ as obtained for each sample are presented in Figs.~1-4. For a better visualisation we extrapolated some fittings toward the value of $T_a(H)$. It is important to note the anomalous enhancement of magnetization appearing above $T_a(H)$ in each curve, which is due to fluctuations of the amplitude of the order parameter with no phase coherence.  Such fluctuations have been recently studied by using the Gaussian GL theory with a \textit{total-energy} cut-off in the fluctuation spectrum to extend its applicability to high reduced temperatures.\cite{EPLVidal} Here we apply this approach to the data above $T_c$ in Figs.~2-4 (see below).

Results of the analysis performed on the $\sqrt{-M}$ vs. $T$ curves (for $H \parallel c$-axis) of YBa$_2$Cu$_3$O$_{6.65}$, and Bi$_2$Sr$_2$CaCu$_2$O$_{8+x}$, produced values of $T_a(H)>T_c$, with $dT_a(H)/dH>0$ which support the scenario of a state with coherent phase persisting above $T_c$ in these systems which are well know to exhibit a pseudo-gap region \cite{timusk}.  Although $T_a(H)$ does not correspond to a true phase transition, it appears to control the scaling properties of the magnetization $M(T)$ in the asymptotic regime, just below the regime of strong amplitude fluctuations. Also, resulting values of the exponent $m$ for YBa$_2$Cu$_3$O$_{6.65}$ are larger than the  mean field value $1/2$.  Resulting values for the exponent $m$ for the Bi$_2$Sr$_2$CaCu$_2$O$_{8+x}$ and YBa$_2$Cu$_3$O$_{6.95}$ crystals were enhanced slightly for $H\parallel \hat{c}$ relative to their mean-field values. Results for YBa$_2$Cu$_3$O$_{6.95}$ crystal with $T_c$ = 92 K confirmed the previous result of Ref.~[\onlinecite{op}] with $T_a(H)\simeq T_c(H)$ obtained in another single crystal with similar doping for both field directions, indicating the absence of phase coherence above $T_c$ for the optimally doped crystal. It should be mentioned that this result is consistent with the fact that optimally doped YBaCuO does not show the pseudo-gap phase \cite{timusk}.   We  mention that results for both YBaCuO samples for the direction $H \parallel ab$ (not shown) also produced values of $T_a(H) \simeq T_c(H)$ in agreement with the mean field behavior expected for $T_c(H)$. A possible explanation for the anomalous behavior of $dT_a/dH$ and $m$ in the underdoped YBaCuO sample as well in Bi$_2$Sr$_2$CaCu$_2$O$_{8+x}$ arises from the $d$-wave pairing symmetry of the order parameter. As discussed in Ref.~[\onlinecite{kwon}], phase fluctuations of a $d$-wave order parameter can have a net effect on the superfluid density of states by reducing the gap in the vicinity of $T_c$.  An inspection of Fig.~2 of Ref.~[\onlinecite{kwon}] shows that the change in the shape of the gap near $T_c$, due to phase fluctuations, is accompanied by a change in the value of the fitting exponent $m$.  In this picture, the dependence of $m$ on $H$ observed for $H \parallel c$ can be explained in terms of the order parameter symmetry which is likely to be $d$-wave in the $ab$-planes of the high-$T_c$ systems.  The presence of nodes or anti-nodes in the order parameter has definite consequences for the phase and amplitude fluctuations \cite{kwon}.  It is important to mention that a value of $m \simeq 1/2$ was found for all curves with $H \parallel ab$ for YBaCuO samples studied here and in Ref.~[\onlinecite{op}]. The observed anisotropy of the exponent $m$ in YBaCuO,  suggests that, in contrast to the effect observed in the gap for the coupling in the $ab$-planes (which is probed when $H \parallel c$-axis), phase fluctuations have a minor effect in the gap for the coupling along the $c$-axis direction, which is probed when $H \parallel ab$-planes. This fact may suggest that the coupling along the $c$-axis direction in YBaCuO has a different symmetry (without nodes) than that in the $ab$-planes, possibly $s$-wave as suggested in Ref.~[\onlinecite{mannhart}] . But, since magnetic field enhance fluctuation effects in high-$T_c$ superconductors \cite{welp}, and YBaCuO is very anisotropic (mainly deoxygenated YBaCuO), experiments with much higher fields for $H \parallel ab$ (which are not available) would be necessary before to conclude for a different symmetry along the $c$-axis direction.

\subsection{Effect of amplitude fluctuations on the $\sqrt{-M}$ vs. $T$ curves above $T_c$}

We discuss below the large amplitude fluctuation observed above $T_a(H)$ in the $\sqrt{-M}$ vs. $T$ curves for all samples. These amplitude fluctuations with no phase coherence, which extend up to temperatures well above $T_a(H)$, can be interpreted in terms of the Gaussian GL theory by introducing a cut-off in the fluctuation spectrum to eliminate the contribution of the high-energy fluctuation modes.\cite{EPLVidal} In the case of 2D high-$T_c$ superconductors it is given by\cite{PC_Carlos}

\begin{eqnarray}
M=-\frac{k_BTN}{\phi_0s}\left[-\frac{c}{2h}\psi\left(\frac{h+c}{2h}\right)-\ln\Gamma\left(\frac{h+\varepsilon}{2h}\right)\right.\nonumber \\
+\left.\ln\Gamma\left(\frac{h+c}{2h}\right)+\frac{\varepsilon}{2h}\psi\left(\frac{h+\varepsilon}{2h}\right)+\frac{c-\varepsilon}{2h}\right].
\label{Prange2D}
\end{eqnarray}
Here $\Gamma$ and $\psi$ are the gamma and digamma functions, $\varepsilon\equiv\ln(T/T_{c})$ the reduced temperature, $N$ the number of superconducting CuO$_2$ layers in their periodicity length, $k_B$ the Boltzmann constant and $c\approx 0.5$ the total-energy cut-off constant.\cite{EPLVidal} In the low magnetic-field limit ($h\ll\varepsilon$) Eq.~(\ref{Prange2D}) may be approximated by
\begin{equation}
M=-\frac{k_BTN}{6\phi_0s}h\left(\frac{1}{\varepsilon}-\frac{1}{c}\right).
\label{Schmidt2D}
\end{equation}
In the case of superconductors moderately anisotropic, the fluctuation magnetization above $T_c$ in the low magnetic field region may be calculated in the Lawrence-Doniach Ginzburg-Landau (GL-LD) approach. In presence of a total-energy cutoff it leads to \cite{EPL_YBCO}
\begin{equation}
M=-\frac{k_BTN}{6\phi_0s}h\left[\frac{1}{\varepsilon}\left(1+\frac{4\xi_c^2(0)}{\varepsilon s^2}\right)^{-1/2}-\frac{1}{c}\right]
\label{Schmidt3D}
\end{equation}
where $\xi_c(0)$ is the coherence length transverse to the CuO$_2$ layers. It is worth noting that in the 2D limit ($\xi_c(0)\ll s$) this expression reduces to Eq.~(\ref{Schmidt2D}), while in the 3D limit ($\xi_c(0)\gg s$) and in absence of a cut-off ($c\to\infty$) it reduces to the well known Schmidt result for 3D materials.\cite{tinkham} 

Due to their high $H_{c2}(0)$ values, the data above $T_c$ for the highly anisotropic YBa$_2$Cu$_3$O$_{6.65}$ and Bi$_2$Sr$_2$CaCu$_2$O$_8$ were analyzed in terms of Eq.~(\ref{Schmidt2D}), while the moderately anisotropic YBa$_2$Cu$_3$O$_{6.95}$ was studied in terms of Eq.~(\ref{Schmidt3D}). As shown in Figs.~2, 3 and 4, the agreement with the data is good, the resulting $\xi_{ab}(0)$ values being 15 \r{A} for YBa$_2$Cu$_3$O$_{6.65}$, 13 \r{A} for Bi$_2$Sr$_2$CaCu$_2$O$_8$ and 6 \r{A} [and $\xi_c(0)=$1 \r{A}] for YBa$_2$Cu$_3$O$_{6.95}$. Taking into account the uncertainty in the fluctuations amplitude (mainly due to the normal-state background, a possible incomplete superconducting fraction,\cite{EPL_f} or the effect of possible $T_c$ inhomogeneities \cite{PRB_Lucia,skelton}), these values are well within the ones in the literature.\cite{tinkham}

\begin{figure}[t] 
\includegraphics[width=\linewidth]{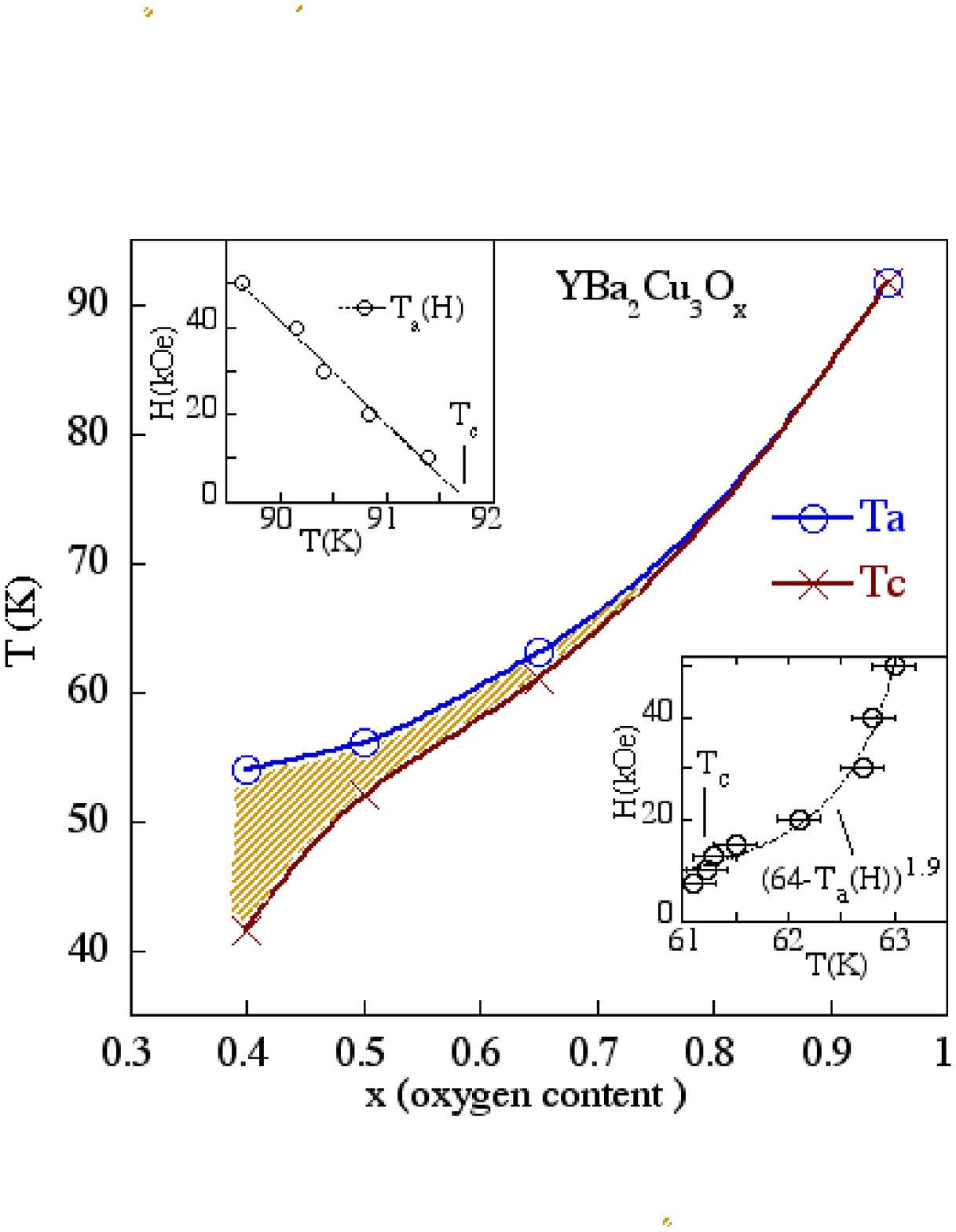}
\caption{Phase diagram for YBaCuO where resulting values of $T_a(H)$ and $T_c$ are plotted as a function of the the oxygen content. The shadowed area represents the region above $T_c$ where phase correlations persist. Top inset: $H$ versus $T_a(H)$ diagram for the optimally doped crystal. Lower inset: $H$ versus $T_a(H)$ diagram for the deoxygenated crystal.}
\label{fig5}
\end{figure} 

\subsection{$T-x$ phase diagram for the superconducting phase fluctuations in YBaCuO}

We finally plot in Fig. 5 a diagram with values of $T_c$ and $T_a(H=50$ kOe) as a function of the oxygen content for  the YBaCuO samples studied here and in Ref.~[\onlinecite{op}]. The lower inset shows values of $T_a(H)$ as obtained for sample with $T_c$= 61.5 K and the top inset shows the same plot for sample with $T_c$=92 K.  As show in the lower inset of Fig. 5, values of $T_a(H)$ for sample with $T_c$=61.5 K obey a power law behavior which tends to a field independent temperature $T_a$=64 K, which is 1 K above the value of $T_a(H)$ for H=50 kOe plotted in Fig. 5. This field independent temperature is probably related to the temperature $T_{\phi}$ defined in Ref.~[\onlinecite{beck}].  For samples with $T_c$= 41 and 52 K the plots of $T_a(H)$ vs. $H$ appearing in Ref.~[\onlinecite{op}] are approximately linear and we also plot in Fig. 5 the corresponding values of $T_a(H)$ obtained for the highest field $H$=50 kOe. It might be possible that this linear behavior of $T_a(H)$ with field as observed for these two samples represents a low-intermediate field region which would give place to a power law behavior (as observed for sample with $T_c$=61.5 K)  for much higher fields. The shadowed region in Fig. 5 evidences the region above the $T_c(H)$ line where phase fluctuations were found to be coherent, which agrees with the existence of a region of temperatures above $T_c$ where phase coherence persists in YBaCuO as predicted in Ref.~[\onlinecite{beck}]. Despite Fig. 5 here was obtained with an applied magnetic field (which gradually probes a temperature region above $T_c$ where phase correlation persists) and Fig. 5 of Ref.~[\onlinecite{beck}] does not consider a magnetic field, it is interesting to compare both figures. A comparison between Fig.~5 here and Fig.~5 of Ref.~[\onlinecite{beck}] suggests that the qualitative behaviour of the $T_a$ line found here is more likely to be related to the $T_1$ line of  Ref.~[\onlinecite{beck}] which values were extracted from specific heat measurements in YBaCuO (see Fig.~2 of Ref.~[\onlinecite{beck}]) rather than with the $T_{\phi}$ line. The $T_{\phi}$ line appearing in Fig.~5 of Ref.~[\onlinecite{beck}] lies well above $T_c$ even for the optimally doped sample, while our results obtained for two distinct optimally doped samples show in this case the onset of phase coherence coinciding with $T_c$. It is important to mention that the values of $T_a(H)$ plotted in Fig.5 are much below the values of the $T_1$ line in Fig.5 of Ref.[\onlinecite{beck}]. Since it appears that the literature lacks on specific heat measurements with field on deoxygenated YBaCuO, it should be considered the possibility that the $T_1$ line in Fig.5 of Ref.~[\onlinecite{beck}] would tend to the $T_{\phi}$ line under an applied magnetic field. For the case of optimally doped ($T_c$=92 K) and slightly underdoped ($T_c$=88 K)YBaCuO crystals the available data in the literature show that values of $T_1$ (the temperature $T_1$ defines a feature in the specific heat occurring above the anomaly as indicated in Ref.~[\onlinecite{beck}]) seem to decrease with field following the shift of the anomaly to lower temperatures as field increases. We mention that the shadowed region in Fig. 5 is probably much larger than it appears, since  we used values of $T_a(H=50 kOe)$ and for samples with $T_c$= 41 and 52~K the values  of $T_a(H)$ are probably much below the values which would result if higher fields were available. 

\section{Conclusions}

In conclusion we studied the effects of phase and amplitude fluctuations in isofield magnetization curves of the high-$T_c$ systems YBa$_2$Cu$_3$O$_{7-\delta}$, and Bi$_2$Sr$_2$CaCu$_2$O$_{8+x}$. The analysis of the asymptotic behavior of $\sqrt{-M}$ with temperature near $T_c$ suggests, with the exception for optimally doped YBaCuO, that all studied systems show phase coherence persisting above $T_c$, where this region above $T_c$ is gradually probed by the magnetic field.  Results of this work allowed to study the evolution of the phase coherence region above $T_c$ in the pseudo-gap region of YBaCuO. The results are in agreement with a theoretical prediction \cite{beck} for high-$T_c$ systems  with a $d$-wave pairing along the $ab$-planes which exhibit a pseudo gap above $T_c$. The large amplitude fluctuation observed for all systems is well explained in terms of uncorrelated fluctuations in a Gaussian Ginzburg-Landau theory with a \textit{total-energy} cut-off in the fluctuation spectrum.

\section*{ACKNOWLEDGMENTS}
We thank H. Beck for  helpfull suggestions and Boyd Veal who kindly provide the YBaCuO  crystals. J.M. acknowledges financial support from the Spanish Ministerio de
Educaci\'on y Ciencia (Grant No. FIS2007-63709), and the Xunta de Galicia
(Grant No.PGIDIT04TMT206002PR). ADA acknowledges support  from CNPq.



\end{document}